# Improving Clinical Diagnosis Performance with Automated X-ray Scan Quality Enhancement Algorithms

Karthik K.[1] and Sowmya Kamath S.[1]

HALE Lab, Department of Information Technology, National Institute of Technology Karnataka Surathkal, Mangaluru - 575025
{2karthik.bhat@gmail.com and sowmyakamath@nitk.edu.in}

**Abstract.** In clinical diagnosis, diagnostic images that are obtained from the scanning devices serve as preliminary evidence for further investigation in the process of delivering quality healthcare. However, often the medical image may contain fault artifacts, introduced due to noise, blur and faulty equipment. The reason for this may be the low-quality or older scanning devices, the test environment or technician's lack of training etc; however, the net result is that the process of fast and reliable diagnosis is hampered. Resolving these issues automatically can have a significant positive impact in a hospital clinical workflow, where often, there is no other way but to work with faulty/older equipment or inadequately qualified radiology technicians. In this paper, automated image quality improvement approaches for adapted and benchmarked for the task of medical image super-resolution. During experimental evaluation on standard open datasets, the observations showed that certain algorithms perform better and show significant improvement in the diagnostic quality of medical scans, thereby enabling better visualization for human diagnostic purposes.

**Keywords:** Image Super-Resolution, Image Enhancement, Image Quality Assessment.

## 1 Introduction

Today by the advancement in medicine, diseases are diagnosed using a variety of diagnostic equipment, among which diagnostic scans play a pivotal role. The earliest clinical predictions are generally obtained via different modalities of medical images such as X-ray images, computerized tomography (CT) and magnetic resonance imaging (MRI) among others [1]. Hence, acquiring good quality diagnostic images is essential for analyzing and determining disease occurrence and progression. Due to the technical restrictions and various economic and physical conditions, there exist several challenges in obtaining a good quality diagnostic image like low resolution (LR) images, under-exposure or over-exposure, occurrence of artifacts introduced by faulty or older scanning equipment etc, which can often render diagnostic scans not suitable for further analysis [2].

Due to this, methods that improve the spatial resolution of medical images are gaining increasing importance in clinical workflow management systems. To create a high resolution (HR) medical image, numerous image enhancement algorithms have been proposed. Image super-resolution (SR) [3] is an approach of restoring high-resolution images from images of lower resolution. Super Resolution can be categorized as Single Image Super Resolution (SISR) and Multi-image Super Resolution (MISR) based on a total of low-resolution images taken as input. SISR can be defined as a method where one low-resolution input image is utilized to restore high-resolution image details. MISR is primarily a reconstruction-based algorithm that takes multiple variants of low-resolution (LR) images and attempts to combine them for recovering the details of the high-resolution (HR) image. Further, enhancing medical images [4] can help medical experts for evaluating diagnosis accurately with more details in pathology research. As a result, medical image enhancement process can substantially enhance the accuracy of computer-aided automatic detection [5].

Even though image super-resolution is an active research field, many medical diagnostic images do not lend themselves well to it. This is due to the inherent structure and manifold information contained in the medical scans. Thus, there is a need for super-resolution techniques to enhance the poor quality images through computational means. This minimizes the overall burden in analysis of the disease for physicians, saving their time and also aiding in accurate diagnosis, specifically in cases where the disease is in its early stages. This also reduces the need for rescanning and wastage of medical resources, and, cost, effort and time of patients, medical personnel and hospital administration.

In this paper, approaches for automatically assessing the quality of diagnostic scan images as part of clinical workflow management is proposed. The exposure level of each input image is analyzed using an exposure-level detection algorithm, and classified as under-exposed, over-exposed or normal, and normalized using image intensity equalization. Ultimately, we improve the image quality using five different algorithms, and report in our observations about their comparative performance using standard metrics. The rest of this paper is structured as follows. Section 2, briefly talks about the existing work of image super-resolution and related image quality assessment mechanisms. Section 3 details about the proposed methodology in improving the quality of X-ray images. Section 4, explores experimental results and observations, followed by conclusion and potential future improvements.

## 2  Related Work

The primary factors that affect medical image quality are noise, edge and contrast. Gaussian noise and impulse noise are two fundamental types that degrade the quality of a medical scan. Generally, median filtering is used to smoothen any impulse noise, but this does not improve the gray-contrast of an image. Histogram Equalization (HE) [6] is one of the widely known method that could be applied to intensify the contrast of the given image; however, the new image that gets developed is often not of acceptable quality. Additionally, grayscale modalities often suffer from low contrast, making the minute details like hairline fractures, fissures etc challenging to identify even for trained medical professionals. Thus, automated radiographic image quality improvement is still an active research work in recent years.

Other factors that affect digital radiographic images are low contrast, visual noise or X-ray scattering and blurring lead by the complexity and density of body tissues.

Radiographic images are often found to need significant improvement in visual quality, including contrast and feature enhancements. Different techniques like Linear Contrast, HE (Histogram Equalization), CLAHE (Contrast limited adaptive histogram equalization) and BPHE (Brightness Preserving Histogram Equalization) were used by Ahmed et al. [7] to enhance digital radiographic images. Georgieva et al. [8] developed an enhancement method for X-ray images using CLAHE followed by morphological processing and noise reduction. CLAHE improves the contrast of an image by reducing the noise in homogeneous areas. But, it was noticed that the artifacts increased when the block size consideration for the image enhancement was more than 16x16. To mitigate this, Yong et al. [9] proposed a hybrid image contrast improvement method formed by the sharp frequency localization-contourlet transform (SFL-CT) and CLAHE. Also, a comprehensive pre-processing algorithm that was developed substantially enhanced the contrast, simultaneously reducing the artifacts.

Obtaining an image with the intended resolution is not easy due to variance in imaging environments and device models. Isaac and Kulkarni [10] proposed a solution to this problem with the use of Super-Resolution (SR) techniques. Their method deals with the dynamic enhancement in the pre-processing step, denoising the medical images and then applying Super-Resolution techniques such as patch-based and orthogonal acquisition algorithms. Huang et al. [11] proposed a two-stage filtering process and contrast enhancement for X-ray images by an adaptive median filter and bilateral filter. The advantage of their method is that it reduced Gaussian and impulsive noise in the image by maintaining the essential structures (i.e., edges) in the images.

Due to advancements in supervised and semi-supervised learning approaches, recent works have focused on the application of Machine learning and Deep Learning models to the problem of medical image enhancement. Zhang and Qiang [12] developed deep learning and transfer learning based super-resolution reconstruction model for enhancing individual low-resolution images. Though significant performance improvement was observed, several limitations still exist, as the computational operations are carried out in a high-resolution space. Specifically, very few works have focused on addressing the problem of over- and under-exposure issues in X-ray images that are often caused by faulty or older imaging devices. These over- and under-exposed images are not suitable for further diagnosis; we aim to effectively improve the quality of such images, with additional emphasis on reducing the computational cost and processing time.

## 3   Proposed Methodology

The overall workflow of the proposed X-ray image quality improvement method is depicted in Fig. 1. The main objective here is to identify, assess and fix faults in the images and improve the visualization so that physicians can appropriately examine the abnormalities. In the first phase, all image exposure levels are checked, based on which, the image is corrected using image intensity equalization method. Different Super Resolution (SR) techniques are employed for improving the image quality for a better visualization, and performance is assessed using standard metrics.

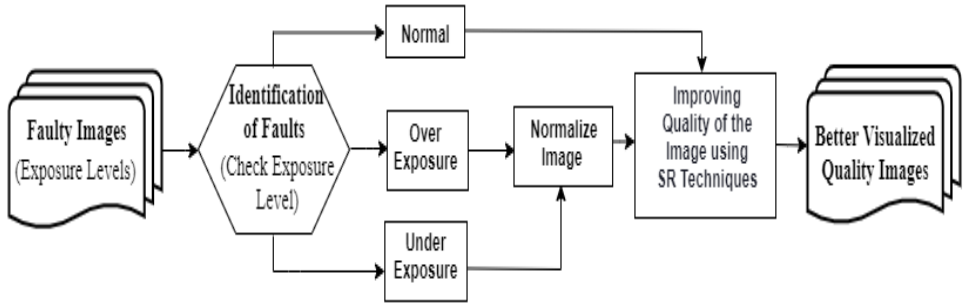

**Fig. 1: Overall Workflow of the proposed X-ray Image Quality Improvement Method.**

The process involved in checking the image exposure level is performed by computing the histogram level of each image. Based on the exposure level an image, can be classified into three categories - under-exposed, over-exposed and normal. A sample of all these types of exposure level images is shown in Fig. 2. To check the image exposure level, we used threshold points considering the image histogram. If the histogram counts are evenly distributed on the scale of 0 to 255, then it is a normal image. If the bin values are more on a scale of 0 to 127, then it is an under-exposed image. Similarly, if the bin values are more on a scale of 128 to 255, then it is an over-exposed image.

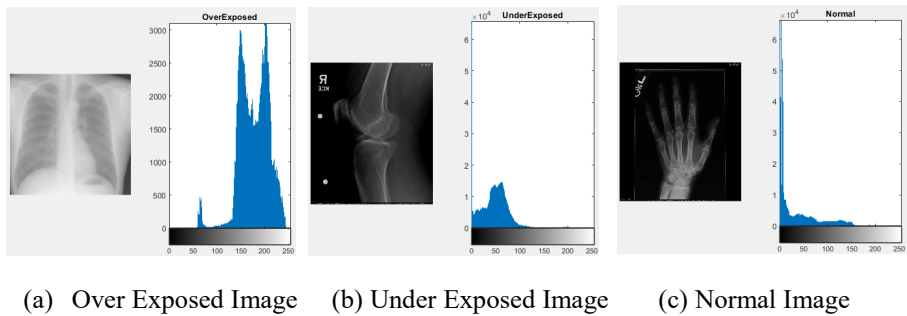

(a) Over Exposed Image   (b) Under Exposed Image   (c) Normal Image

**Fig. 2: Images with different exposure levels and Histogram.**

If an image is over or under-exposed, then the image is normalized using image intensity equalization, as detailed earlier. In the next step, the normalized images are enhanced using enhancement algorithms like Contrast Limited Adaptive Histogram Equalization (CLAHE) and Unsharp Masking (UM) using the Gaussian filtering technique. CLAHE is a histogram-based method used to improve contrast in images, which computes the histogram for the region around each pixel in the image, improving the local contrast and enhancing the edges in each region. Adaptive Histogram Equalization (AHE) over amplifies the noise in the image; CLAHE prevents this by limiting the amplification.

To apply CLAHE to the images, we first convert them to grayscale and then normalize. This approach is similar to N-CLAHE, but we do not use log normalization. The implementation of CLAHE requires three inputs - *window size* (the size of the rectangular region around the pixel to be processed), *clip limit* (maximum number of occurrences of the pixel in the histogram) and *iterations* (Number of clipping iterations). After this step, the image is padded by reflecting the pixels in the borders. Then, for each pixel in the image, we calculate the clipped histogram for the region around it, i.e., we define the maximum number of occurrences a pixel should have. If the occurrence is greater than the clip limit, we cut the exceeding area and redistribute to all other pixels. To improve the technique, this process can be repeated a certain number of times until we get a desired contrast image. With this clipped histogram, we calculate the probability of each pixel in it and compute the CDF (Cumulative Distribution Function), using the cumulative sum of the ordered pixels, and multiply each value of the function by 255, to limit the image's values to [0, 255]. After calculating the CDF, all pixels will have a transformation value. We now apply this transformation to the pixel in the center of the region. We noticed that, for certain images, the image becomes very noisy when the clip limit chosen is very high. This may be because, when the limit is very high, no clipping is performed, and the CLAHE algorithm is essentially similar to AHE algorithm.

Unsharp masking is a linear filter that is capable of amplifying high frequencies of an image. The first step of the algorithm is to copy the original image and apply a Gaussian blur into it (Blur intensity is defined by a setting called Radius). If we deduct the blurred image from the original image, we will obtain only the edges created by the blur, which is called the unsharped mask. The radius setting is related to the blur intensity because it defines the size of the edges. The amount, on the other hand, controls the intensity of the edges (how much dark or light it will be). The experimental results, observations and further developments are discussed in the subsequent sections. Finally, the enhanced image is collected after computed and visualized using Eq. (1).

$$sharpened\_image = original\_image + amount * (unsharped\_mask) \quad (1)$$

We also employed bicubic interpolation for upscaling the low-resolution(LR) image, that resulted to a high-resolution image where the dimension is similar to that of the reference image. Further, neural network SR methods like Very-Deep Super-Resolution (VDSR) and Single image super-resolution CNN were utilized in developing a high-resolution(HR) image. VDSR network [13] builds a HR image using a single LR image by learning and mapping the difference in its frequency. The network consists of an image input layer, then with a 2-D convolutional layer that consist of 64 filters. A total of 20 convolutional layers, each of which that follows a ReLU activation layer builds up the network, which introduces nonlinearity in the network. A image patch size of 41-by-41 is used and the network was trained for 100 epochs deploying stochastic gradient descent with momentum (SGDM) optimization. The learning rate was initially fixed to 0.1 and decreased with a factor of 10 for each 10 epochs. VDSR works with a surplus learning strategy, i.e., the network learns to assess with a surplus image. A surplus image informs regarding the high-frequency characteristics of an image. For super-resolution cases, a surplus image is a variation with a HR reference image and a LR image that has upscaled using bicubic interpolation by matching the dimension of the image to the reference image.

SRCNN [14] learns pixel mapping between LR and HR images with pre-processing optimization techniques. The first phase deals with patch extraction and representation, where, patches ($f_1 X f_1$) from LR image are extracted and each patch is represented as a HD vector, which is a set of feature maps equivalent to its dimensions. The next operation is non-linear mapping, which performs a non-linear mapping of each HD vector ($n_1$) to another HD vector ($n_2$). Here, each mapped vector represents a HR patch. Finally, reconstruction operation combines all the HR patches ($f_2 X f_2$) to generate the final HR image. We set the parameters as $f_1 = 9$, $f_2 = 5$, $n_1 = 64$ and $n_2 = 32$ while constructing HR image from a LR image. The learning rate is set to $10^{-4}$ in initial two layers, and $10^{-5}$ in the final layer. Empirically, a small learning rate was set to the last layer for the network to converge. The main importance of SRCNN is that it attained good results compared to the state-of-the-art methods.

## 4  Experimental Results

For experimental purpose, a small dataset of medical images from MedPix[1] consisting of 66 different body organ images was used. Standard evaluation metrics like PSNR (Peak Signal-to-Noise Ratio) and SSIM (Structural Similarity Index) were used for measuring the image quality, i.e. quality of the reconstructed image. Given two images, $I$ (ground truth image) and $\hat{I}$ (reconstructed image), both of same size, the MSE and the PSNR (in dB) is given by Eq. (2) and (3), where, $I$ is the maximum intensity of a grayscale image i.e. 256. $X_{ij}$ and $Y_{ij}$ are the intensity of original and reconstructed image. MSE is the Mean Square Error, M and N are the number of rows and columns in the image. SSIM [15] is a visual metric that measures quality of reconstructed image based on the effect of luminance, contrast and structure (as per Eq. (4), where, $C_1$ and $C_2$ are constants used to keep away the uncertainty when $\mu_x$ and $\mu_y$ are very close to zero. $\sigma_x, \sigma_y$ are contrast comparison functions).

$$\text{PSNR} = 10 \log_{10} \left[ \frac{I^2}{\text{MSE}} \right] \quad (2)$$

$$\text{MSE} = \frac{1}{[N \times M]^2} \sum_{i=1}^{N} \sum_{j=1}^{M} (X_{ij} - Y_{ij})^2 \quad (3)$$

$$\text{SSIM}(\mathbf{x}, \mathbf{y}) = \frac{(2\mu_x \mu_y + C_1)(2\sigma_{xy} + C_2)}{\left(\mu_x^2 + \mu_y^2 + C_1\right)\left(\sigma_x^2 + \sigma_y^2 + C_2\right)} \quad (4)$$

Evaluation of the five X-ray image quality improvement methods on a sample of three images is shown in Fig. 3. A simple GUI also was developed for illustrating the quality enhancement by visualizing the original and corrected image, as shown in Fig. 4. Based on our observations, we found that the UM and the CLAHE methods performed well in improving the quality of the image for a better visualization, whereas, in terms of HR image display from a single LR image SRCNN outperformed among other SR methods by achieving good PSNR/SSIM index values.

---
[1] The National Library of Medicine MedPix®, **https://medpix.nlm.nih.gov/home**

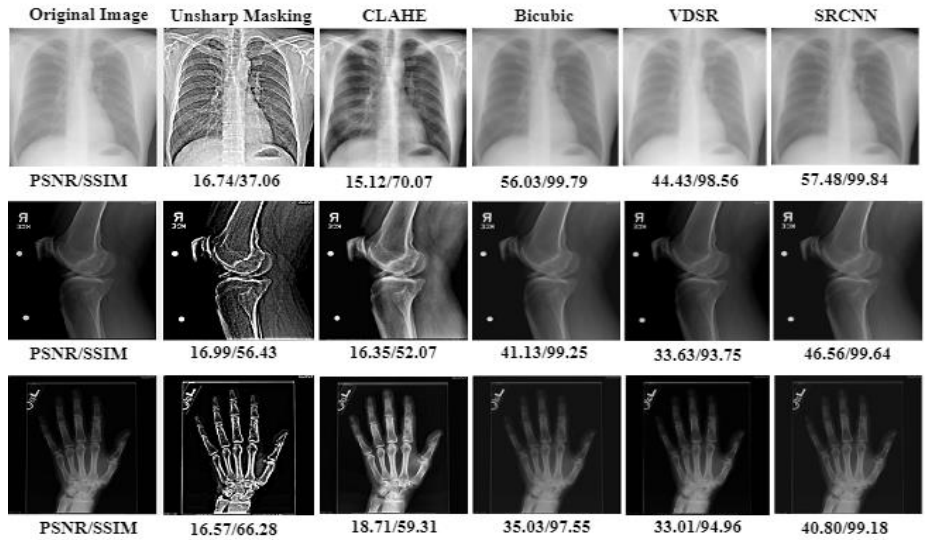

Fig. 3: Comparative Evaluation of UM, CLAHE, Bicubic, VDSR and SRCNN for X-ray image enhancement.

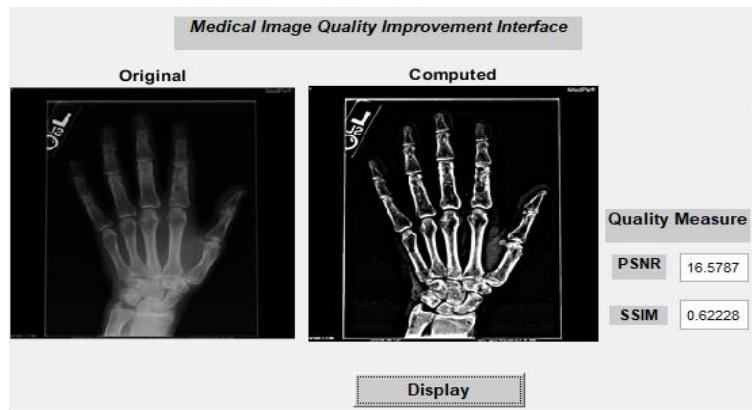

Fig. 4: Illustration of X-ray Image Enhancement with quality metrics.

## 5  Conclusion & Future Work

In this paper, five image super-resolution algorithms - Unsharp mask using gaussian filter, CLAHE, Bicubic Interpolation, VDSR network and SRCNN model were implemented for image super-resolution, and evaluated for better visualization of X-ray images. Experiments were performed to comparatively evaluate these 5 approaches. Based on the visualized enhanced images, it can be observed that the processed image captured hidden features well through edge and contrast enhancement, in turn amplifying the visibility of region of interest or artifacts. A patch size of 16 pixels (4X4) in Bicubic interpolation resulted in a smoother image, while VDSR showed better performance while transforming a LR image to HR. SRCNN outperformed all other methods due to its lightweight architecture and superior learning behavior. As another part for future task, our plan is to experiment with GAN based neural models for medical image enhancement. We also aim to design a real-time application, considering the need for achieving good computation time along with good performance.

**Acknowledgment:** The authors gratefully acknowledge the financial support provided by the DST-SERB Early Career Research Grant (ECR/2017/001056) and the facilities at the Department of Information Technology, NITK Surathkal.